\documentclass[twocolumn,showpacs,amsmath,amssymb,aps,prd,floats,nofootinbib]{revtex4-2}
\usepackage{graphicx}
\usepackage[switch,pagewise,running]{lineno}
\usepackage{dcolumn}
\usepackage{bm}
\usepackage{epsfig}
\usepackage{xspace}
\usepackage{hyperref}
\hypersetup{colorlinks,linkcolor=blue,citecolor=blue,urlcolor=blue}  
\usepackage{color,xcolor}
\usepackage{verbatim}

\begin{document}
\title{Detection of the cosmological evolution of the Doppler factor for blazars jets
}
\author{Dahai Yan$^{1}$, Haiyun Zhang$^{1}$, Jianeng Zhou$^2$, and Lijuan Dong$^{1}$}

\affiliation{$^1$Department of Astronomy, Key Laboratory of Astroparticle Physics of Yunnan Province, Yunnan University, Kunming 650091, People's Republic of China\\
$^2$Shanghai Astronomical Observatory, Chinese Academy of Sciences, 80 Nandan Road, Shanghai 200030, People's Republic of China\\
}
\begin{abstract}
The Doppler factor ($\delta$) is a fundamental quality for the relativistic jets from  
active galactic nuclei (AGNs).
It is also fundamental to assessing the number of the entire population of the jetted
AGN at high redshift ($z$), and therefore is important for tracing the growth of supermassive black holes (SMBHs) across cosmic time.
Here we present the identification of the positive cosmic evolution of the Doppler factor 
in {\it Fermi}-detected bright $\gamma-$ray blazars.
The redshift dependence of the Doppler factor, $\delta\propto(1+z)^{0.8}$, 
is measured from the observed characteristic energies in the gamma-ray spectra of 141 blazars.
Moreover, the analysis of the characteristic timescales derived from modeling the long-term optical light curves of 89 blazars with Gaussian process regression 
gives $\delta\propto(1+z)^{1.1}$, but with a larger scatter.
Note that each index is derived from the entire sample, representing an average evolution. 
Interestingly, the index itself also appears to evolve, with low-luminosity sources showing stronger evolution, as indicated by a larger index.
This detection straightly suggests that relativistic jets from AGNs are 
much more common at high redshifts than what is previously estimated.
\end{abstract}

\maketitle

\section{Introduction} \label{sec:intro}
Active galactic nuclei (AGNs) can form relativistic jets that are observed throughout the entire
electromagnetic spectrum from radio wavelengths to TeV energies.
The production mechanism of the relativistic jet is still unknown.
The current evidence suggests that the relativistic jet's formation 
may be related to the supermassive black hole spins and the magnetization of the accretion disk 
\citep[see][for a recent review]{2019ARA&A..57..467B}. 
High-resolution radio observations enable us to gain insights from the extragalactic relativistic jet on the scale of astronomical units \citep[e.g.,][]{2017A&ARv..25....4B}.
High-precision multiwavelength/messenger campaigns are improving our knowledge of the AGN jets.

One important problem concerning the AGN jet is its cosmological evolution \citep[e.g.,][]{2012ApJ...751..108A,2019MNRAS.490..758Q,2017ApJ...846...78Y}.
Observations of blazars make them the best choice for studying the cosmic evolution of the relativistic jet.
Blazar is believed to be a unique tool 
for inferring the number of the entire jetted AGNs at high redshift \citep[e.g.,][]{2011MNRAS.416..216V,2013MNRAS.432.2818G,2015MNRAS.446.2483S,2022A&A...660A..74B}.
A blazar is an AGN with a relativistic jet directed nearly to Earth.
Relativistic beaming of electromagnetic radiation from the jet makes blazars appear 
much brighter than the intrinsic emissions.
This relativistic beaming effect also makes the timescale for variability 
shorter than the intrinsic quantity and boosts photons to higher energies.
The relativistic beaming effect is quantified by the Doppler factor ($\delta$) which is defined as
$\delta=[\Gamma(1-\beta \rm cos\theta)]^{-1}$
where $\beta$ is the bulk velocity of the jet matter in units of the speed of light, 
$\Gamma=(1-\beta^2)^{-1/2}$ is the corresponding Lorentz factor, and $\theta$ is the angle between the velocity vector and the line of sight.
Observed radiation of blazar mainly comes from the non-thermal radiation of relativistic jets.
The observed jet radiation indeed contains information about the Doppler factor.

Blazars have been known to be variable in the entire electromagnetic spectrum \citep[e.g.,][]{2017A&ARv..25....2P}.
The blazar long-term light curves detected at optical and GeV energies are accurately presented as a damped random walk (DRW) with various damping timescales ($\tau_{\rm DRW}$) \citep{2022ApJ...930..157Z}.
These timescales for a blazar sample range from a few days to two hundred days \citep{2023ApJ...944..103Z}.
This timescale includes information on the Doppler factor of the jet.

The observations of {\it Fermi} make precise measurements 
for the energy spectra of thousands of blazars over the energies from 100 MeV to more than 300 GeV  \citep{2020ApJ...892..105A}.
The results demonstrate all bright blazars have
curved spectra in the LAT energy range, which is quantified as a log-parabolic function $N(E)\propto(E/E_b)^{-\alpha -b\cdot{\rm log}(E/E_b)}$ 
where $\alpha$ is the spectral index at $\sim E_b$, 
$b$ is the curvature parameter, and $E_b$ is a characteristic energy of the spectrum 
where the spectral behavior transitions from a nearly power-law form to a curved or parabolic form \citep[e.g.,][]{2020ApJ...892..105A}.
Because of the relativistic beaming effect, the emission is blue-shifted by a factor of $\delta$.

These observed characteristic quantities are dependent on the Doppler factor.
When the blazar redshift is available, one can measure the redshift dependence of the Doppler factor for blazars.
In this work, we study the cosmological evolution of the blazar jet's Doppler factor by analyzing the long-term optical light curves and the GeV gamma-ray spectra.

\section{Methodology and dataset} \label{sec:style}
We consider two independent quantities, $\tau_{\rm DRW}$ and $E_b$, 
which is measured from the long-term optical light curve and GeV spectrum respectively.
We have the following relations between the observed and the intrinsic quantities,
$\tau_{\rm DRW}=\tau^{\rm int}_{\rm DRW}(1+z)/\delta$ and $E_b=E^{\rm int}_b\delta/(1+z)$ where $\tau^{\rm int}$ and $E^{\rm int}$ refer to the intrinsic quantities and $z$ is the blazar redshift.
The term $(1+z)$ represents the impact of the expanding space on the two quantities.

Here we consider the redshift dependence of $\delta$ to be of the form of $(1+z)^m$.
Replacing $\delta$ with $\delta(1+z)^m$, we rewrite the above relations, i.e.,
\begin{equation}
\tau_{\rm DRW}=\frac{\tau^{\rm int}_{\rm DRW}}{\delta}(1+z)^{1-m},\  \\
E_b=E^{\rm int}_b\delta(1+z)^{m-1}.
\label{eq1}
\end{equation}
Clearly, $m=0$ demonstrates no redshift dependence of the Doppler factor.
We treat $m$ as a free parameter and determine it from the observations.

To achieve our aim, we first need a blazar sample with reliable redshifts, 
and the redshift should have a broad dynamic range.
We choose the blazar sample presented in \citet{2014Natur.515..376G}.
Most sources in this sample are $\gamma$-ray-bright flat spectrum
radio quasars (FSRQs) .

The {\it Fermi}-LAT data of each blazar covering from August 2008 to August 2023 in this sample are analyzed with the standard procedure by using FermiTools v2.0.8. In detail, 0.1-300 GeV PASS 8 data within a region of interest of $20^\circ \times 20^\circ$ centered at the position of each blazar are used, and events with zentih angle $>90^\circ$ are excluded to minimize the contamination from the Earth limb. 
The background in likelihood analysis is modeled by the sources cataloged in an incremental version of the fourth full catalog of LAT sources (4FGL-DR4; \cite{2022ApJS..260...53A,2023arXiv230712546B}) and the diffuse models (gll\_iem\_v07 for the Galactic diffuse and iso\_P8R3\_SOURCE\_V3\_v1 for the extragalactic isotropic diffuse respectively). The 15-year average energy spectrum for each blazar is constructed.
We adopt the log-parabolic function to fit the {\it Fermi} gamma-ray spectrum to determine $E_b$.
Finally, we have 141 blazars with well-measured $E_b$, and $E_b\approx 0.3-3\ $GeV.
The redshifts of the 141 blazars cover the range $z=0.065-3.033$.

\citet{2023ApJ...944..103Z} has built high-quality optical light curves for 38 blazars using the photometric data from the SMARTS and Steward Observatory (SO) monitoring projects. 
\citet{2023ApJ...944..103Z} adopted a Gaussian process (GP) regression \citep{2017AJ....154..220F} to 
model these optical light curves, and found that a DRW kernel\footnote{It is defined by an
exponential covariance function $k(t_{nm})=2\sigma^2_{\rm DRW}\cdot {\rm exp}(-t_{nm}/\tau_{\rm DRW})$, where $t_{nm}=|t_n-t_m|$ is the time lag between measurements $m$ and $n$, $\sigma_{\rm DRW}$ is the amplitude term, and $\tau_{\rm DRW}$ is the damping timescale. } 
can accurately describe each blazar light curve.
The characteristic timescale (i.e., the damping timescale in DRW model) for the blazar variability is calculated. 
More details on applying GP regression to AGN variabilities (including quasar accretion disk variability and blazar jet variability) are presented in \citet{2021Sci...373..789B} and \citet{2022ApJ...930..157Z,2023ApJ...944..103Z}.
Here we use the data of blazars from Zwicky Transient Facility (ZTF) program to expand the sample of optical light curves in \citet{2023ApJ...944..103Z}.
The DRW kernel in the GP method is used to fit each light curve, and then to calculate the timescale $\tau_{\rm DRW}$. The timescale is required to be larger than the mean cadence of the light curve and smaller than one-tenth of the light curve's length, ensuring the calculated timescale is reliable.
Finally, we have 89 blazars with reliable $\tau_{\rm DRW}$, and $\tau_{\rm DRW}\approx9-210\ $days.
The redshifts of the 89 blazars are adopted from \citet{2014Natur.515..376G} and \citet{2023ApJ...944..103Z}  covering the range $z=0.03-2.54$.

\section{Results} \label{sec:floats}

We have two samples from which we can extract two independent characteristic quantities. 
One sample is used to obtain the well-measured $\gamma$-ray $E_b$, 
and the other one is used to measure the reliable optical $\tau_{\rm DRW}$.
The two independent observations are used to determine the cosmological evolution index $m$ for the Doppler factor of the blazar jet.
From Eq.(\ref{eq1}), we have
\begin{equation}
\begin{aligned}
 {\rm log_{10}}\tau_{\rm DRW}={\rm log_{10}} (\frac{\tau^{\rm int}_{\rm DRW}}{\delta})+(1-m) {\rm log_{10}}(1+z),\ \   \\
 {\rm log_{10}} E_b={\rm log_{10}} (E^{\rm int}_b\delta)+(m-1) {\rm log_{10}}(1+z).
\label{eq2}
\end{aligned}
\end{equation}
After taking the logarithm, they are two simple linear relations.
A hierarchical Bayesian approach is used to perform linear regression.
The regression assumes $y=b\pm(m-1)*x+\sigma$ where $b$ is a normalization term, $x$ is ${\rm log_{10}}(1+z)$
, and $\sigma$ presents the intrinsic random scatter about the regression.

\begin{figure}
    \centering
    \includegraphics[width=1\linewidth]{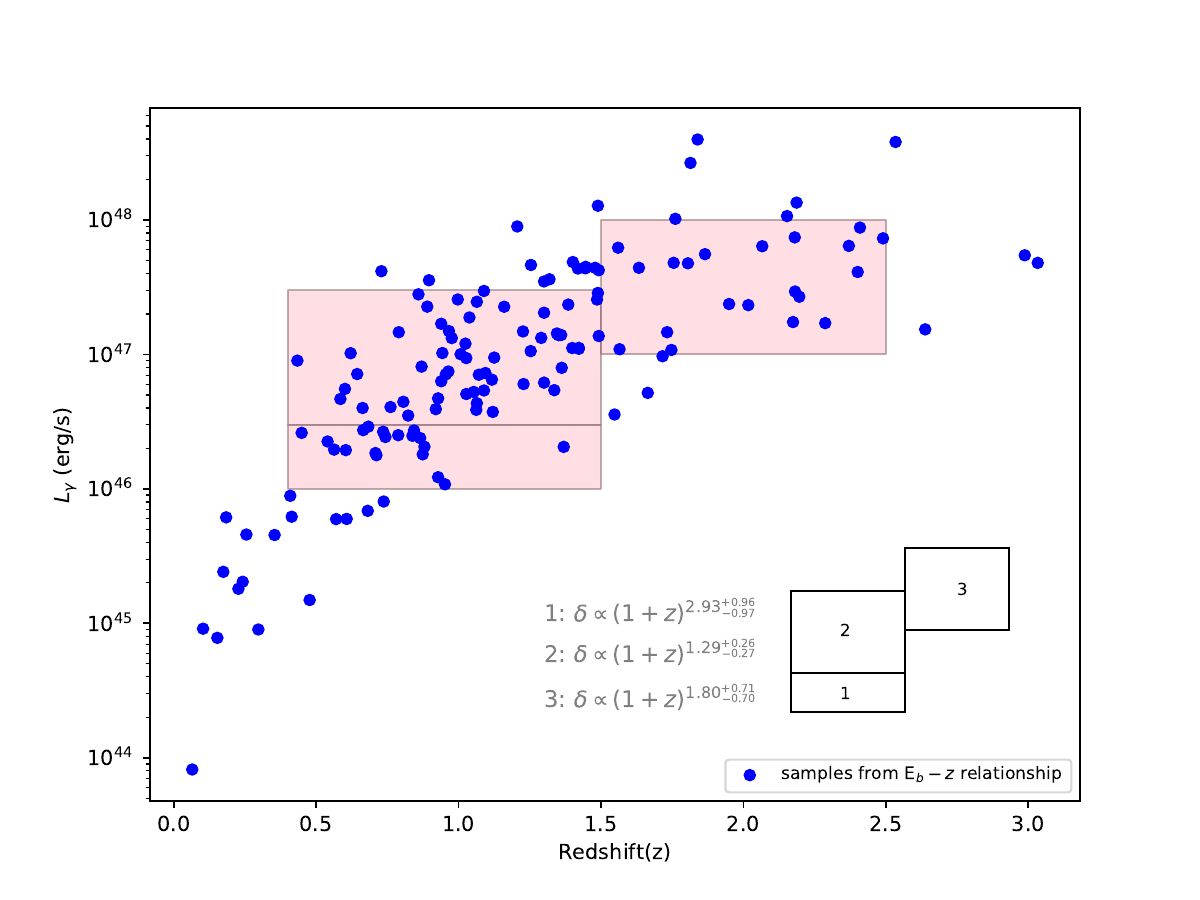}
    \caption{Blazar sample as a function of redshift and $\gamma$-ray luminosity. 
    The sample is used to measure the quantity of $E_b$.  
    Three subsamples (the rectangles) are considered to test the impact of the select effect on the redshift dependence of the Doppler factor. }
    \label{fig:L1}
\end{figure}

\begin{figure*}
    \centering
    \includegraphics[width=0.45\linewidth]{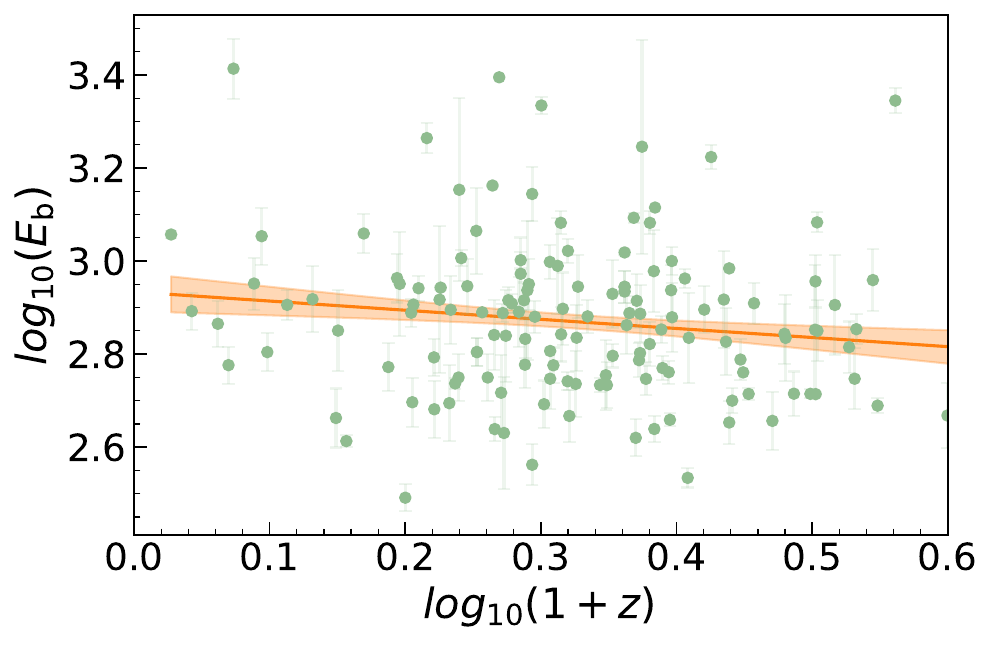}
    \includegraphics[width=0.45\linewidth]{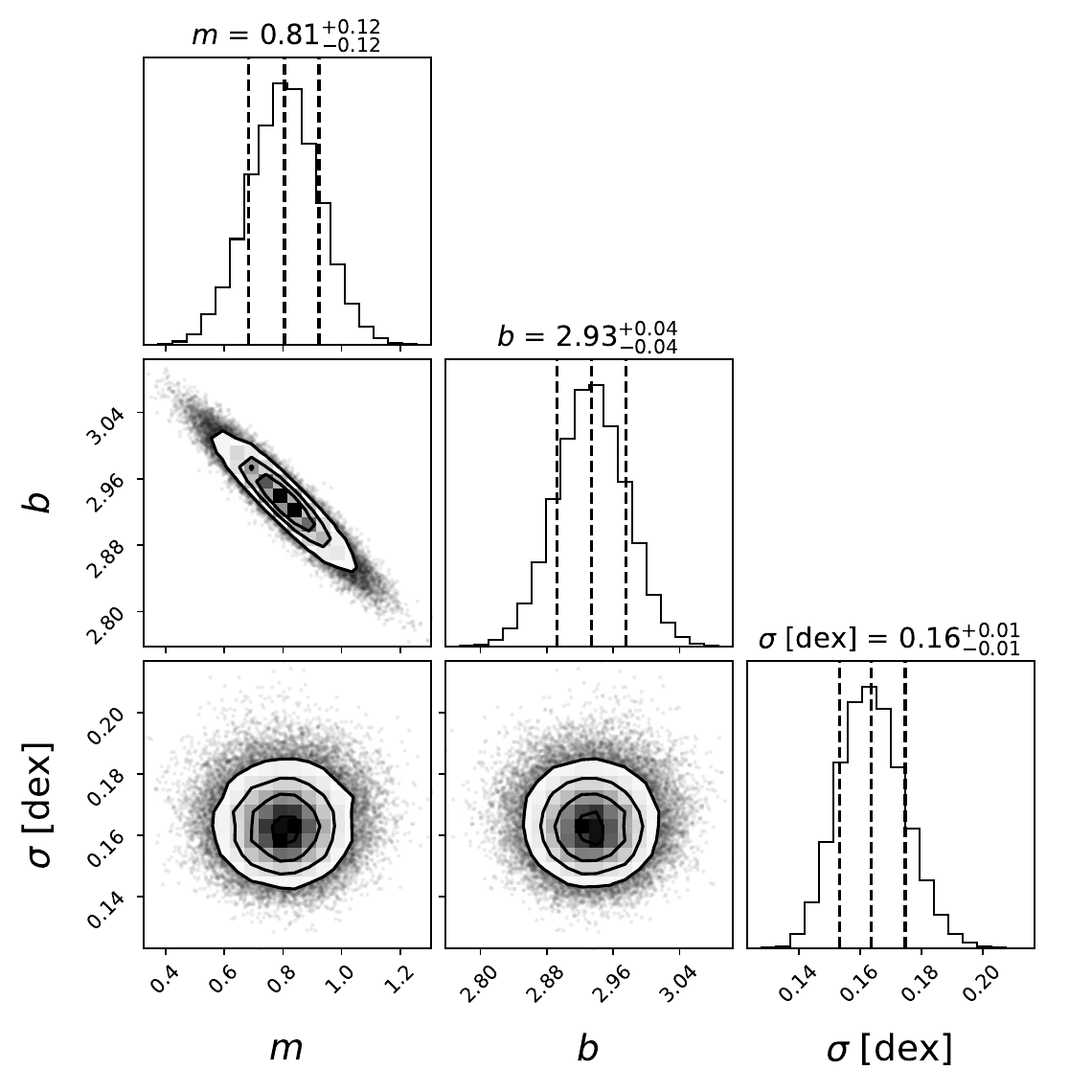}
    \caption{Fitting results for the entire spectral sample. Left: Linear regression result for the characteristic energies; Right: Posterior distributions for the parameters. }
    \label{fig:energy}
\end{figure*}

Figure~\ref{fig:L1}  displays the $\gamma$-ray luminosity plotted
against redshift for the spectral sample. For the entire sample, one can see a dependence of the $\gamma$-ray luminosity on the redshift.
First the entire sample is used to determine the cosmological evolution index $m$ for the Doppler factor of the blazar jet.
The fitting results are presented in Figure~\ref{fig:energy}.
The value of $m=0.81\pm0.12$ is clearly offset from zero, 
indicative of a redshift dependence of the Doppler factor in the blazar sample.

In order to test the impact of the selection effect, 
we consider three subsamples (see Figure~\ref{fig:L1} ).
No clear dependence of the $\gamma$-ray luminosity on the redshift is presented in each of the three subsamples.
The positive cosmological evolution index $m$ obtained from each subsample significantly differs from zero.
The values of $m$ are reported in Figure~\ref{fig:L1}.
Furthermore, the index itself appears to be evolving.
Compared with the high-luminosity blazars ( subsamples 2 and 3),
the relatively low-luminosity blazars ( subsample 1) have a stronger evolution with $m\approx2-4$.

\begin{figure}
    \centering
    \includegraphics[width=1\linewidth]{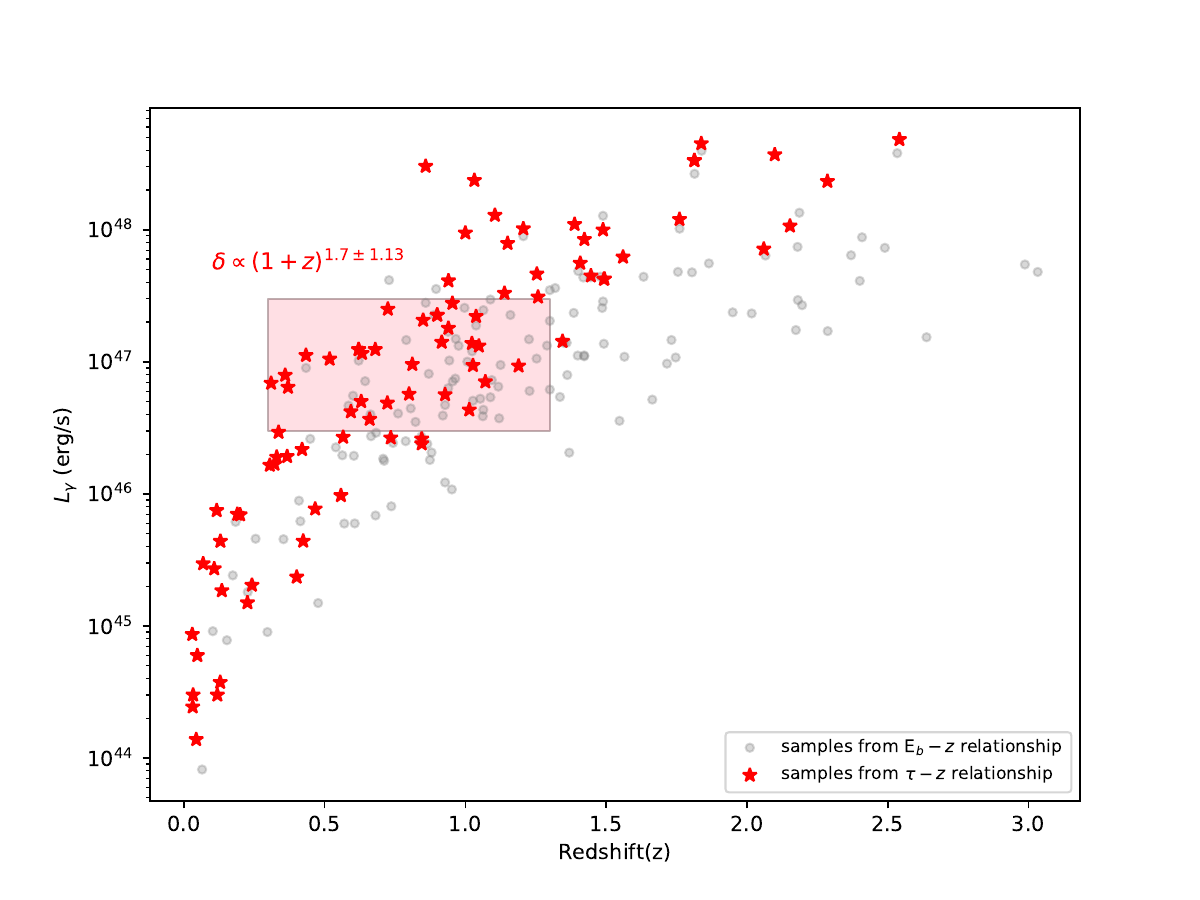}
    \caption{Blazar sample as a function of redshift and $\gamma$-ray luminosity. 
    The grey dots represent the spectral sample, and the red stars the variability sample.
   For  the variability sample, one subsample (the rectangle) is considered to test the impact of the select effect on the redshift dependence of the Doppler factor. }
    \label{fig:L2}
\end{figure}

\begin{figure*}
    \centering
    \includegraphics[width=0.45\linewidth]{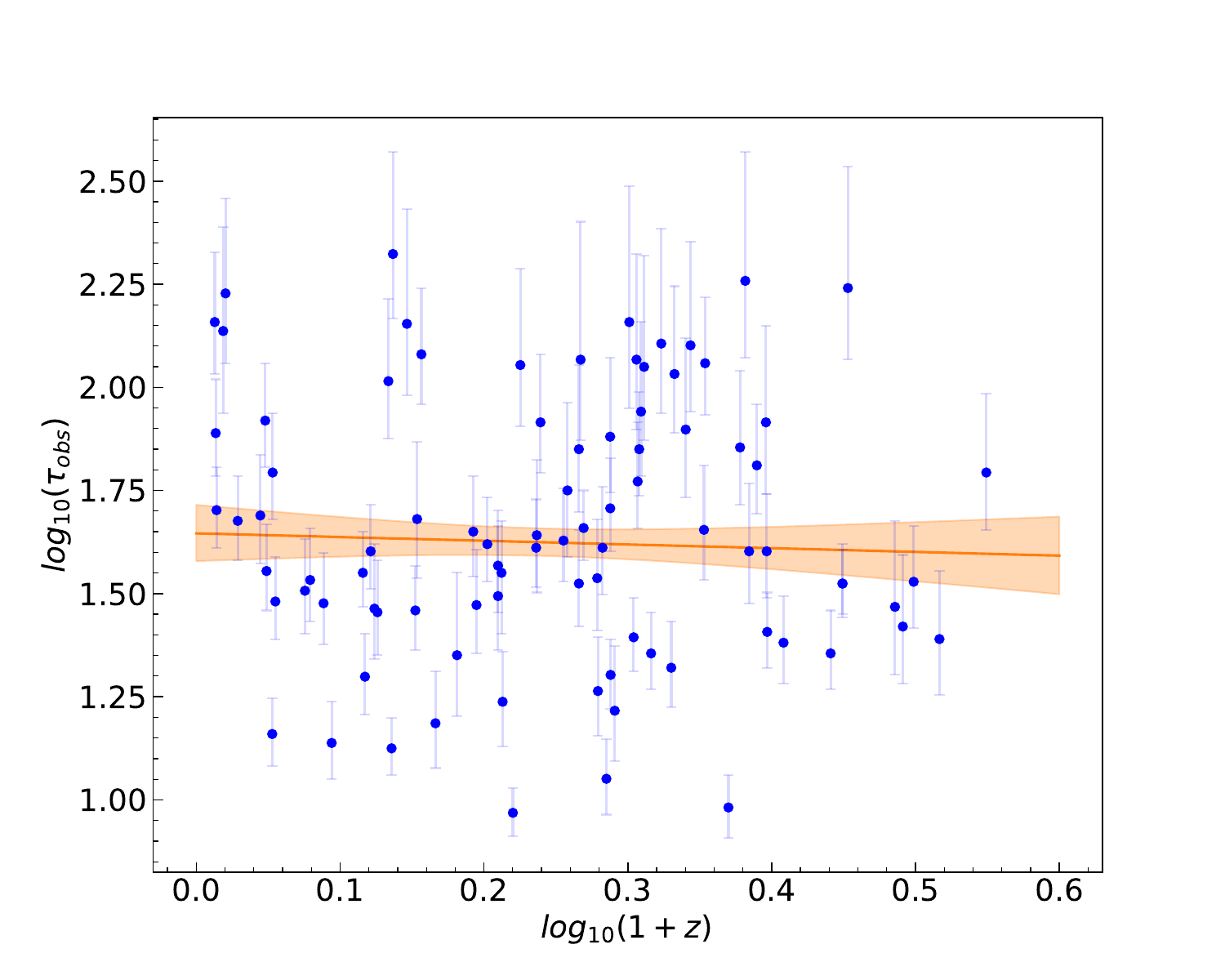}
    \includegraphics[width=0.45\linewidth]{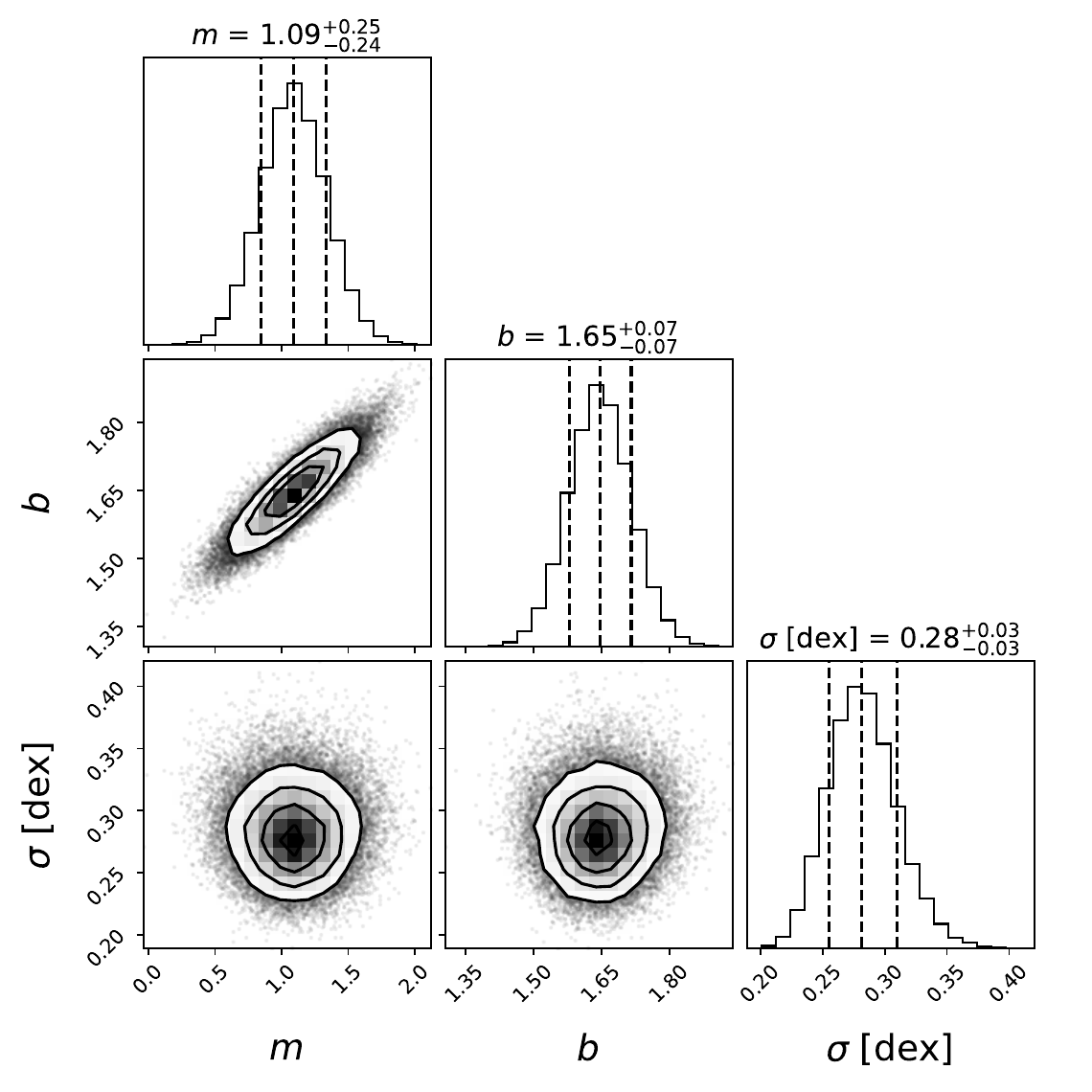}
    \caption{Fitting results for the entire variability sample. Left: Linear regression result for the characteristic variability timescales; Right: Posterior distributions for the parameters.}
    \label{fig:variability}
\end{figure*}

Figure~\ref{fig:L2}  displays the $\gamma$-ray luminosity plotted
against redshift for the variability sample, together with the spectral sample.
In general, the variability sample has a higher $\gamma$-ray luminosity than the spectral sample.
The results obtained from the entire variability observations are presented in Figure~\ref{fig:variability}.
The result of $m=1.09^{+0.25}_{-0.24}$ independently confirms the redshift dependence of the Doppler factor.

Due to the small number of blazars in the variability sample, only one subsample is considered to test the impact of the selection effect.
Again,  we require that the sources in this subsample should be as evenly distributed as possible in the $L_{\gamma}-z$ plot.
For this subsample, the obtained $m$ exhibits larger uncertainties but still differs significantly from zero.

\section{Conclusions and Discussion} \label{sec:cd}

In this paper, we report the detection of the redshift dependence of the Doppler factor from the blazar jet emissions,
i.e., $\delta\propto(1+z)^m$ with $m\approx0.8-1$ on average.
This redshift dependence of the Doppler factor is found in blazars with the redshift up to $\sim3$.
This finding demonstrates that a high-$z$ blazar has a larger Doppler factor than 
a low-$z$ blazar.
Taking $\delta\sim10$ for low-$z$ blazars, 
the redshift dependence of the Doppler factor expects $\delta\sim30-40$ for a blazar with redshift $z=3$.
This is consistent with the implication of the 
fast variabilities with the timescale of a few hours observed in gamma-ray blazars with redshift $z\geq3$ \citep{2018ApJ...853..159L,2019ApJ...879L...9L}.

Blazars with small Doppler boosting factors are rarely observed at high redshifts. 
Indeed, \citet{2021ApJ...923...67H} demonstrated a clear positive correlation between $\gamma$-ray luminosity and the Doppler factor. 
To evaluate whether our results are affected by this observational selection effect, 
we analyze several subsamples in the $L_{\gamma}-z$ plot (Figures~\ref{fig:L1} and \ref{fig:L2}), 
ensuring that the sources within each subsample are as evenly distributed as possible. 
This simple method helps reduce the influence of selection bias.
In every subsample, we find a positive $m$ that significantly deviates from zero (Figures~\ref{fig:L1} and \ref{fig:L2}), 
supporting the validity of our result. 
Additionally, the cosmological evolution of the Doppler factor appears to be complex, 
with low-luminosity sources exhibiting stronger evolutionary trends. 
However, further exploration is currently limited by the small size of the sample.

The number of blazars with $\theta\leq1/\Gamma$ is related to the total jetted AGN population (including the obscured AGNs),
$N_{\rm jetted}=2\Gamma^2N_{\rm blazar}$ \citep[e.g.,][]{2024arXiv240707236B,2022MNRAS.511.5436D}.
This makes blazars a unique tool for inferring the number density of jetted AGNs at high redshifts, free from
obscuration effects.
In the previous studies, $\Gamma$ is considered to be independent of redshifts.
The redshift dependence of the Doppler factor found here arises from the cosmological evolution of $\Gamma$,
i.e., $\Gamma\propto(1+z)^{0.8-1}$. This cosmic evolution has a great influence on the census of the total population of the high-$z$ jetted AGN. 
At $z=3$, the comoving space density of blazars $n_{\rm blazar}$ is estimated to be roughly one per $\rm Gpc^3$ \citep{2022MNRAS.511.5436D}. 
Based on this value and the redshift dependence of $\Gamma$, we expect about 290 jetted AGNs per $\rm Gpc^3$ at $z=3$, assuming $\Gamma=4$ for low-$z$ blazars.
This number is about the space density of the entire AGN population at $z=3$ calculated by \citet{2022MNRAS.511.5436D} using the bolometric quasar luminosity function (QLF) given in \cite{2020MNRAS.495.3252S}.
The jetted fraction (the ratio of the jetted AGN's number to the total AGN's) is estimated to be 10-20 percent in the local Universe, 
and it is still in the range of 10\%-20\% at $z=3$ \citep[e.g.,][]{2022MNRAS.511.5436D}.
If this jetted fraction is true, 
our result indicates that the current QLF \cite[e.g.,][]{2020MNRAS.495.3252S}  underestimates the number of the total AGN population.
This implies that more AGNs could be in an obscured phase, as the previous studies suggested \citep[e.g.,][]{2011MNRAS.416..216V,2024arXiv240707236B,2024ApJ...961L..25L}.

The detection of the redshift dependence of the Doppler factor
directly suggests that relativistic jets from AGNs are more common at high redshifts 
compared to what is usually estimated \citep[e.g.,][]{2022A&A...663A.147S}, 
consistent with the implication of the results derived from radio-detected AGN samples by \citet{2024A&A...684A..98C}. Our finding could be useful for understanding the role of the relativistic jets in the growth of the SMBHs at high redshifts.

At last, it’s worth noting that while our approach reduces the impact of selection effects, 
it could not entirely eliminate them. 
Regardless, the current data indicate that the evolution of Dopper factor of blazar jets over cosmic time is indeed real.


\acknowledgements
D.H.Yan acknowledges funding support from
the National Natural Science Foundation of China (NSFC)
under grant No. 12393852.
H.Y.Zhang's work is supported by the State-sponsored Postdoctoral Researcher program GZB20230618. J.N.Zhou is supported by NSFC (grant U2031205) and Youth Innovation Promotion Association CAS.

\appendix

\bibliography{sample631}{}
\bibliographystyle{aasjournal}
\end{document}